# The synergistic modulation of electronic and geometry structures leads to ultra-low thermal conductivity of graphene-like borides (g-$B_3X_5$, X=N, P, As)

Linfeng Yu[1], Jinyuan Xu[1], Chen Shen[2], E Zhou[1], Jing Wu[1], Hongbin Zhang[2], Xiong Zheng[1], Huiming Wang[3], and Guangzhao Qin[1,*]

[1]*State Key Laboratory of Advanced Design and Manufacturing for Vehicle Body, College of Mechanical and Vehicle Engineering, Hunan University, Changsha 410082, P. R. China*

[2]*Institute of Materials Science, Technical University of Darmstadt, Darmstadt 64287, Germany.*

[3] *Hunan Key Laboratory for Micro-Nano Energy Materials & Device and School of Physics and Optoelectronics, Xiangtan University, Xiangtan 411105, Hunan, China*

## Abstract

The design of novel devices with specific technical interests through modulating structural properties and bonding characteristics promotes the vigorous development of materials informatics. Herein, we propose a synergy strategy of component reconstruction by combining geometric configuration and bonding characteristics. With the synergy strategy, we designed a novel two-dimensional (2D) graphene-like borides, *e.g.* g-$B_3N_5$, which possesses counter-intuitive ultra-low thermal conductivity of 21.08 W/mK despite the small atomic mass. The ultra-low thermal conductivity is attributed to the synergy effect of electronics and geometry on thermal transport due to the combining reconstruction of g-BN and nitrogene. With the synergy effect, the dominant acoustic branches are strongly softened, and the scattering absorption and Umklapp process are simultaneously suppressed. Thus, the thermal conductivity is significantly lowered. To verify the component reconstruction strategy, we further constructed g-$B_3P_5$ and g-$B_3As_5$, and uncovered the ultra-low thermal conductivity of 2.50 and 1.85 W/mK, respectively. The synergy effect and the designed ultra-low thermal conductivity materials with lightweight atomic mass cater to the demand for light development of momentum machinery and heat protection, such as aerospace vehicles, high-speed rail, automobiles.

* Author to whom all correspondence should be addressed. E-Mail: gzqin@hnu.edu.cn

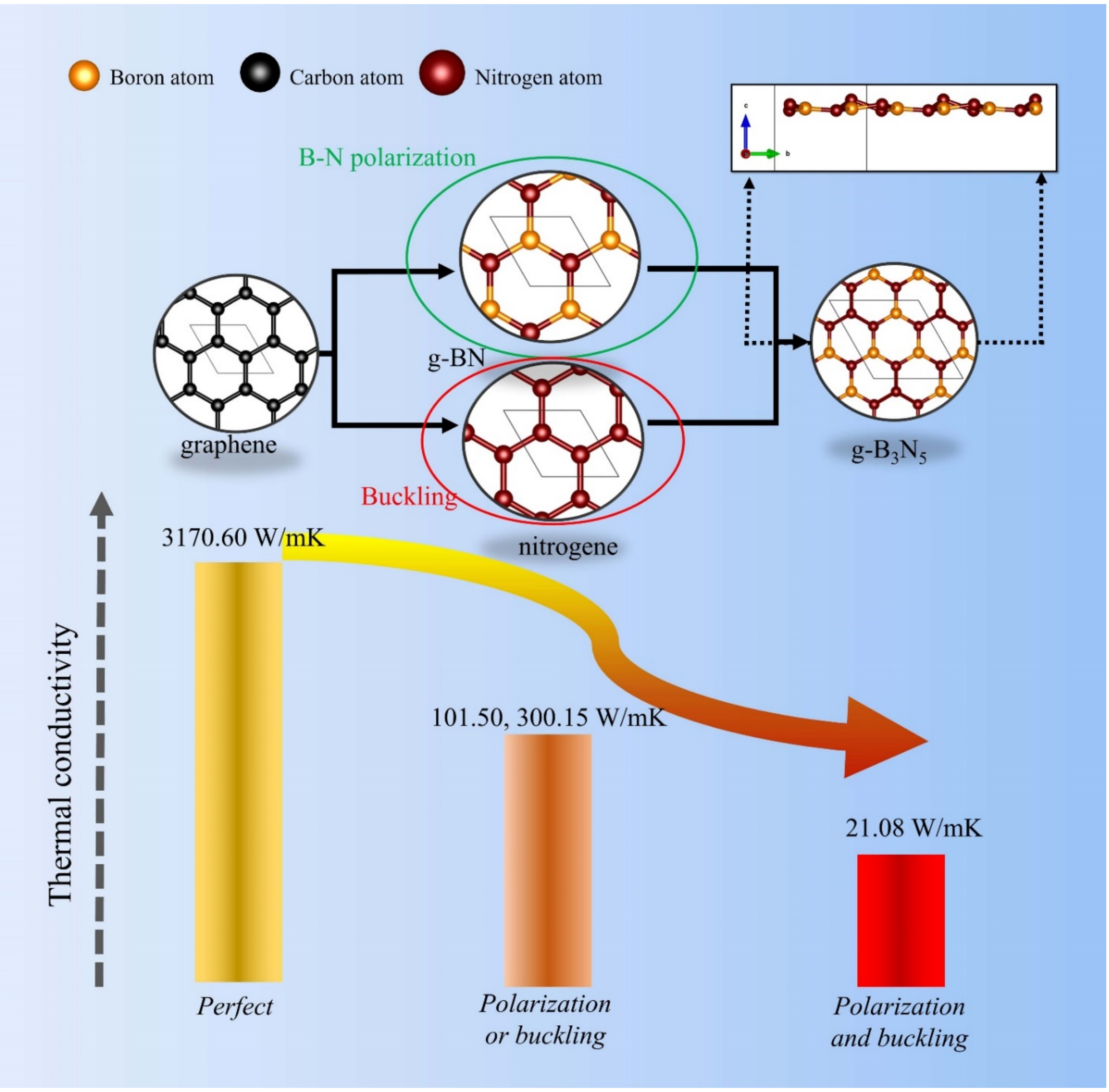

## 1. Introduction

The structural and bonding characteristics of materials and other specific features that endue their physical and chemical properties are essential for searching novel systems with specific technological interests. In thermal science, several conceptions have been proposed to search materials with low $\kappa$, such as the lone-pair electrons[1,2], the resonance bond[3–5], the rattling model,[6–8]and the quasi-bond model[9]. The features are highly representative in characterizing strong anharmonicity, which have been widely studied to achieve fundamental insights into thermal transport from the view of electronic structure.

From the valence shell electron pair repulsion theory, the model of the lone-pair electron plays an extremely vital role in chemistry and physics, which is widely used in various situations[10–13]. The lone electrons pair refers to a pair of non-bonded *s* electrons that have solid stereochemical activity and interact with adjacent bonding electrons to produce nonlinear electrostatic repulsion. The essence behind lone-pair electrons is that the anharmonic interatomic forces are contributed by overlapping variable wave functions during thermal vibration[1,2]. By comparing the orbital evolution of $AgInTe_2$ and $AgSbTe_2$[14], the phenomenon of the lone-pair electrons can be visually understood. For instance, the energy of the $5s^2$ orbital of the In atoms in $AgInTe_2$ is similar to that of the $5p^1$ orbital, and thus the outermost electrons tend to participate in bonding. In contrast, the 5*s* orbital electrons of the Sb atoms in $AgSbTe_2$ are significantly lower than the 5*p* orbital in energy, which forms a typical feature of the lone-pair electrons from the band structure. The lone-pair electrons generally exist in certain nitrogen group elements, such as $As^{3+}$, $Sb^{3+}$, and $Bi^{3+}$, which will lead to strong anharmonicity. The lone-pair electrons model provides a clear and intuitive physical image for understanding the fundamentals of thermal transport.

However, previous researches on lone-pair electrons mainly focused on three-dimensional (3D) systems[1,14–17]. Recently, we have extended the concept to two-dimensional (2D) materials, such as graphene-like XN (X=B, Al, and Ga), which possess lower thermal conductivity relative to bulk materials[18–22]. Interestingly, different from g-GaN (14.9 W/mK[23]), the g-BN has a higher $\kappa$ over two hundred despite the presence of lone-pair electrons, which is due to its lighter atomic mass and planar structure. In addition, low $\kappa$ of 3D nitrides are rarely reported in literature, owing to the light atoms usually lead to a high $\kappa$. Thus, graphene-like materials with light-atom how to achieve the low $\kappa$ when the lone-pair electrons strategy is not practical?

In the field of thermal management, high thermal conductivity ($\kappa$) can benefit the applications in heat dissipation. Besides, materials with low $\kappa$ are also urgently needed due to their potential applications in thermal barrier coatings[24,25], rewritable storage devices[26,27] thermoelectric materials[28], *etc*. Most importantly, materials with lightweight atoms and low $\kappa$ have thermal insulation applications in kinetic energy equipment with lightweight requirements, such as automobiles,[29,30] aerospace vehicles[31,32]. However, the thermal conductivity is negatively correlated with the atomic mass. Conventionally, low $\kappa$ tends to be concentrated in heavy mass regions while light mass contributes to a high $\kappa$. Thus, it is challenge to achieve extremely low $\kappa$ in the light materials.

In this work, a novel 2D boron nitrogen compound g-$B_3N_5$ is constructed using g-BN and nitrogene based on the component reorganization strategy. Noted that the combination of electrons and geometric structures (lone pair electrons and buckling structure) forms a new graphene-like boron nitrogen configuration, which has light average atomic mass but ultra-low $\kappa$ (21.08 W/mK) compared with graphene. With this synergy strategy, we further designed g-$B_3P_5$ and g-$B_3As_5$ with ultra-low $\kappa$ of 2.50 and 1.85 W/mK, respectively, based on the reconstruction of blue phosphorene and g-BP (or arsenene and g-BAs).

# 2. Results and discussion

## 2.1 Light atoms and ultra-low thermal conductivity

Commonly, the thermal conductivity is negatively correlated with the atomic mass. Fig. 1 presents the $\kappa$ of 2D graphene-like materials as a function of average atomic mass. Conventionally, low $\kappa$ tends to be concentrated in heavy mass regions (fourth quadrant) while light mass contributes to high $\kappa$ (first quadrant). Light atoms are represented by boron, nitrogen, and carbon elements. For instance, graphene is believed to have the highest $\kappa$ of 3000-5000 W/mK[33–35] among all the known materials. In addition, both experimental and theoretical studies have reported that graphene-like boron nitrogen (g-BN) has a high $\kappa$ of 227-625 W/mK[36–40]. As the average atomic mass increases, the $\kappa$ of materials tends to decrease. When the $\kappa$ is lower than 100 W/mK, the average atomic mass of materials is mostly larger than 20, and they are generally composed of elements behind the third period. For instance, the $\kappa$ of silicene (19 W/mK[19]) and blue phosphorene (23 W/mK[41]) with heavy atoms is lower than that of nitrogene (101.05 W/mK) although they have the same buckling structure. In short, materials with low $\kappa$ tend to accumulate in the area of heavy atoms, as shown in the fourth quadrant of Fig. 1. Thus, it is challenge to achieve extremely low $\kappa$ in the light materials.

For graphene, the combination of perfect planar structure and strong covalent bonding lead to its ultra-high $\kappa$[19]. In g-BN, the emerging lone-pair electrons bring strong anharmonicity to reduce $\kappa$[42], despite the similar planar structure and average atomic mass as graphene. Different from the g-BN, the low $\kappa$ of nitrogene originates from the buckling structure, similar to silicene[43,44]. As for the g-$B_3N_5$, the anomalously ultra-low thermal conductivity as shown in Fig. 1 can be attributed to the synergy effect of buckling structure and lone-pair electrons as discussed following.

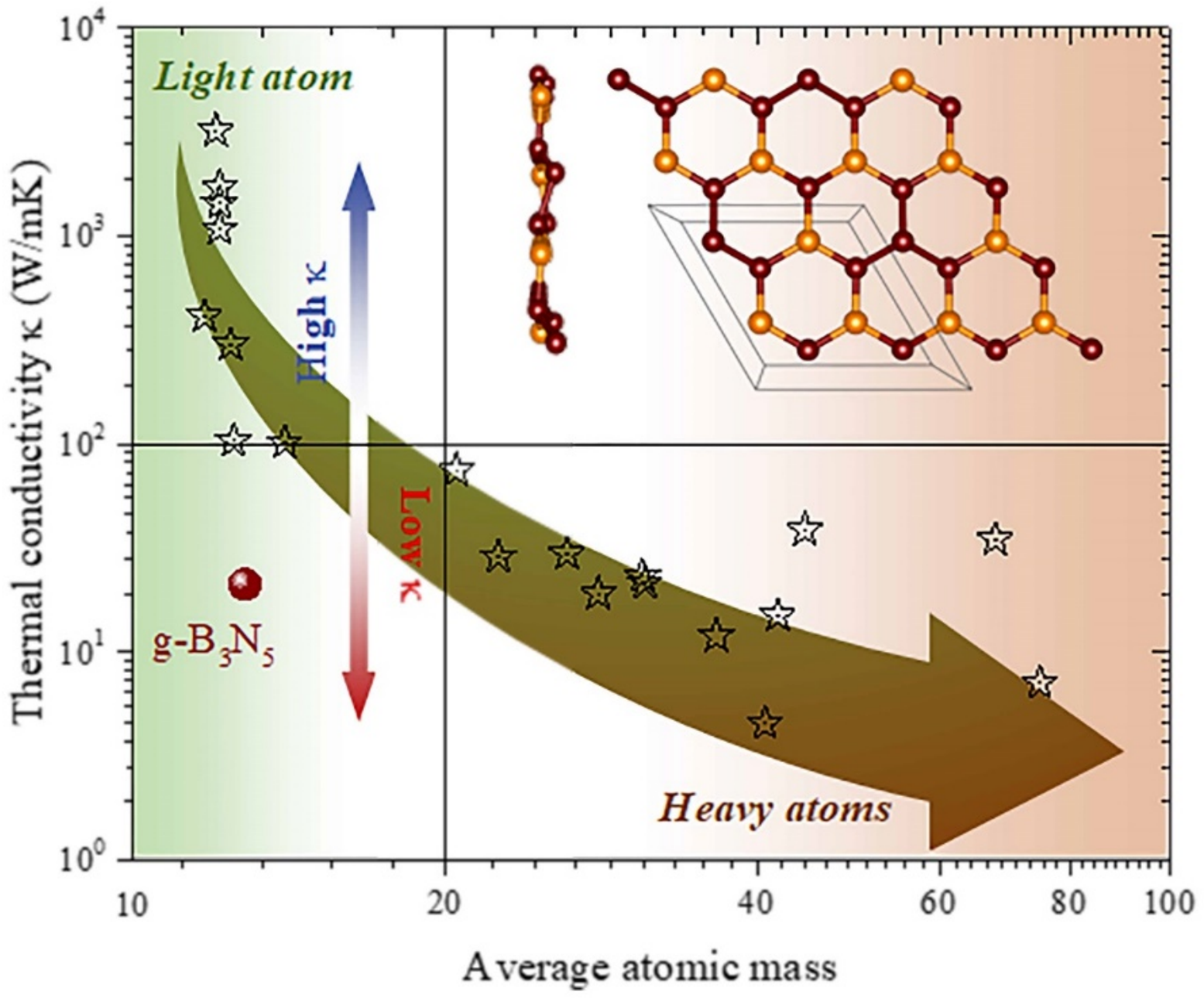

Fig. 1. Thermal conductivity as a function of average atomic mass at 300 K for 2D graphene-like materials[18,23,34,41,45–47], where the data are listed in the Supplementary Table S1. The insert is the top and side views of the geometric structure of g-$B_3N_5$. The solid black lines at the average atomic mass of 20 and the thermal conductivity of 100 W/mK divide the figure into four quadrants. The $\kappa$ of 2D materials with atomic mass lower than 20 is mainly concentrated in the (top-left) second quadrant, which tend to have $\kappa$ larger than 100 W/mK. In contrast, materials with low thermal conductivity tend to be in the fourth quadrant due to the heavy atomic mass.

## 2.2 The component reorganization and phonon softening

Compared with graphene and g-BN, the two orders of magnitude drop in the $\kappa$ of g-$B_3N_5$ can be attributed to the synergy effect of electronics and geometry on thermal transport due to the component reconstruction of g-BN and nitrogene. Fig. 2(a) presents a sketch map of the top-bottom recombination process of g-$B_3N_5$. The original components with four atoms of triangular coordination come from graphene. Furthermore, boron (B) and nitrogen (N) atoms are used to respectively replace the center and vertex carbon (C) atoms to form the g-BN's essential component, which possesses triangular plane coordination with the lone-pair electrons. Simultaneously, the triangular coordinate carbon atoms in graphene can also be replaced by N atoms to construct a buckling structure, constituting the essential component of nitrogene. Ultimately, g-$B_3N_5$ is formed by combining the triangular elements of g-BN and nitrogene and sharing the nitrogen atoms of the vertex.

Phonon dispersion is of great significance in indicating the dynamic stability and evaluating the lattice $\kappa$. The phonon dispersion of g-$B_3N_5$ along the high symmetry point Γ-M-K-Γ are plotted in Fig. 2(b). More information about phonon dispersion is shown in Supplementary Note S1 and Fig. S1. The highest vibration frequency of the optical mode in g-$B_3N_5$ at point Γ is 42.0 THz, which is close to graphene (47.1 THz), g-BN (40.27 THz), and nitrogene (32.23 THz). The corresponding total density of states (DOS) of phonons also reveals the same high-frequency vibration range due to their lighter atomic mass. However, the highest frequency of the acoustic branch in g-$B_3N_5$ (7.3 THz) shows a significant reduction compared to graphene (29.9 THz), g-BN (26.4 THz), and nitrogene (24.4 THz), which indicates the strong softening of the acoustic phonon branches as shown in Fig. 2(c). The softened acoustics branches will contribute to reduced phonon group velocity and enhanced the phonon-phonon scattering[5,48]. The FA and TA branches of g-$B_3N_5$ and g-BN did not form the Dirac cone at point K, while the Dirac cone appears in graphene and nitrogene. It seems that the polarization bond can destroy the phonon Dirac cone. In addition, the FA and TA branches of g-$B_3N_5$ display the phenomenon of avoid-crossing. These negative phenomena in the phonon dispersion significantly reduced the $\kappa$ of g-$B_3N_5$[49,50].

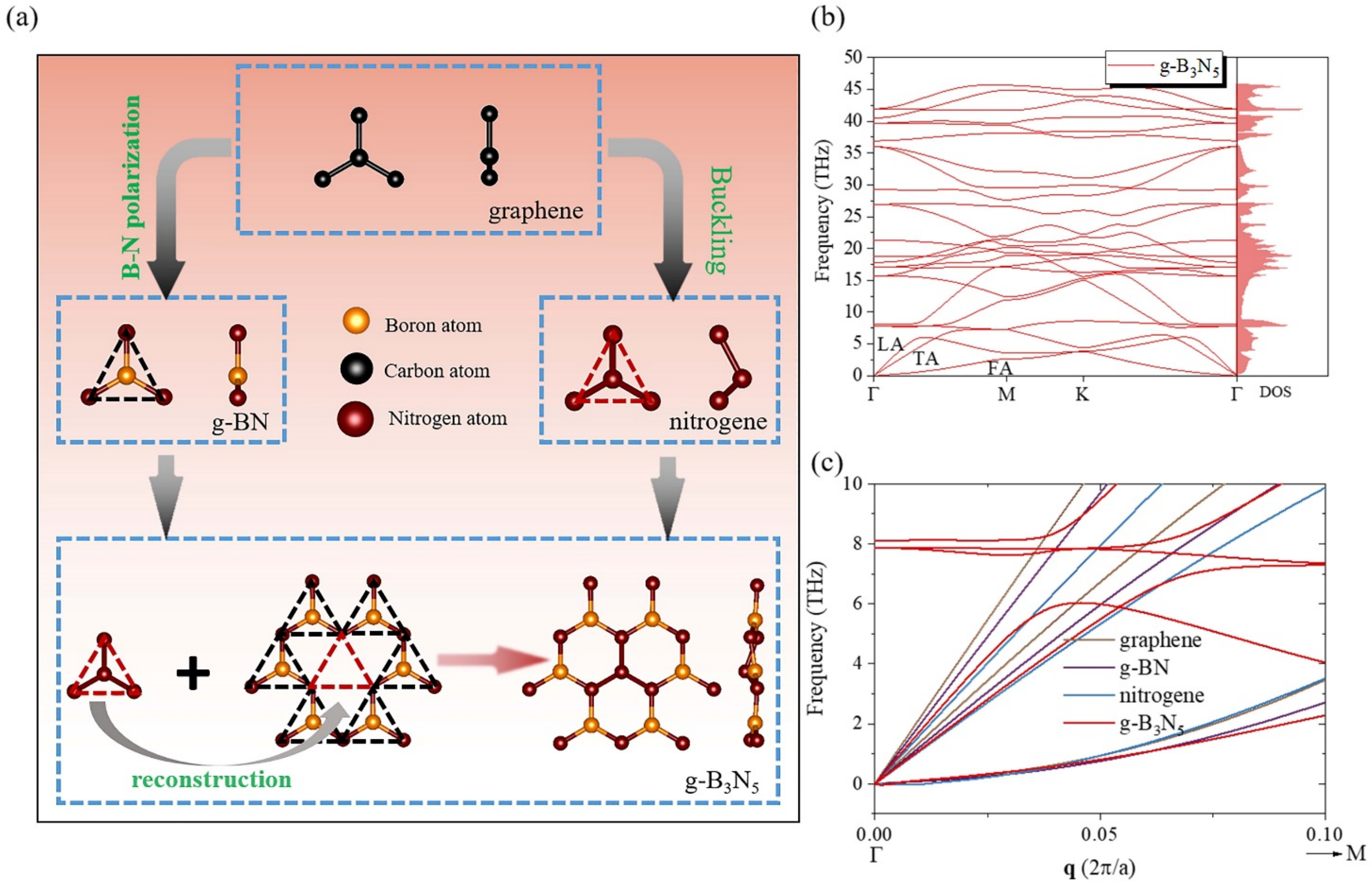

Figure 2. (a) The work flow of the component reconstruction strategy with g-$B_3N_5$ as an example (graphene→(g-BN & nitrogene)→g-$B_3N_5$). (b) The phonon dispersion and phonon density of states (DOS) for g-$B_3N_5$. (c) Softened acoustic phonon branches along the high symmetry points (Γ → M).

## 2.3 The synergistic effect of geometry and electronic structure

Fundamental physical insights can be further obtained based on the evolution of electronic structures. Fig. 3(a-f) provide the evolution of the electronic density of states during geometric reorganization. The highest $\kappa$ of graphene can be attributed to the combination of strong C-C bonding and planar structure[18,19,51]. As shown in Fig. 3(a), *s* electrons and *p* electrons in graphene are hybridized in the valence band region, and thus the electrons gather between C atoms to form a strong C-C bond with $sp^2$ hybridization. But the *s* electrons of the N atom in g-BN are much lower in energy than the *p* electrons [Fig. 3(d)], forming a lone pair of *s* electrons. The lone-pair electrons forbid the $sp^2$ hybridization in the N atom while remaining in the B atom [Fig. 3(c)]. This asymmetric orbital evolution originates from the B-N polarization due to the difference in electronegativities like g-AlN[18] and g-GaN[19,23], contributing to stronger anharmonicity than graphene. As for nitrogene, the buckling structure weakens the $sp^2$ hybridization, causing deviation of the *s* electrons from the bonding region in Fig. 3(b). Compared with the planar structure, the buckling structure breaks mirror symmetry, leading to the centrosymmetric distribution of the electron cloud, which enhances the nonlinear Coulomb repulsion between the electrons. Thus, the $\kappa$ of nitrogene is also one order of magnitude lower than that of graphene.

Though the component recombination, g-$B_3N_5$ has the negative factors both of g-BN (the lone-pair electrons) and nitrogene (buckling structure), which leads to a lower κ of two orders of magnitude than graphene. In g-$B_3N_5$, the *s* electrons of the N atom deviates further from the bonding region when the *p* electrons mainly contribute to the bonding region. The deviated *s* electrons are isolated and localized at different non-bonding energy levels [Fig. 3 (f)] while the B atoms remain $sp^2$ hybridized [Fig. 3 (e)] after the reconstruction. The evolution of orbital asymmetry can be intuitively understood based on the analysis of electron localization function (ELF) as shown in Fig. 3 (g). Unlike graphene, nitrogene, and g-BN, the asymmetric atomic configuration of g-$B_3N_5$ leads to uneven distribution of bond length and bond angle. The distorted geometric environment will lead to asymmetrical evolution of electron distribution as the isosurface of 3D ELF in Fig 3 (h), which more easily contributes to anharmonic atomic forces between the N and B atoms during thermal vibration.

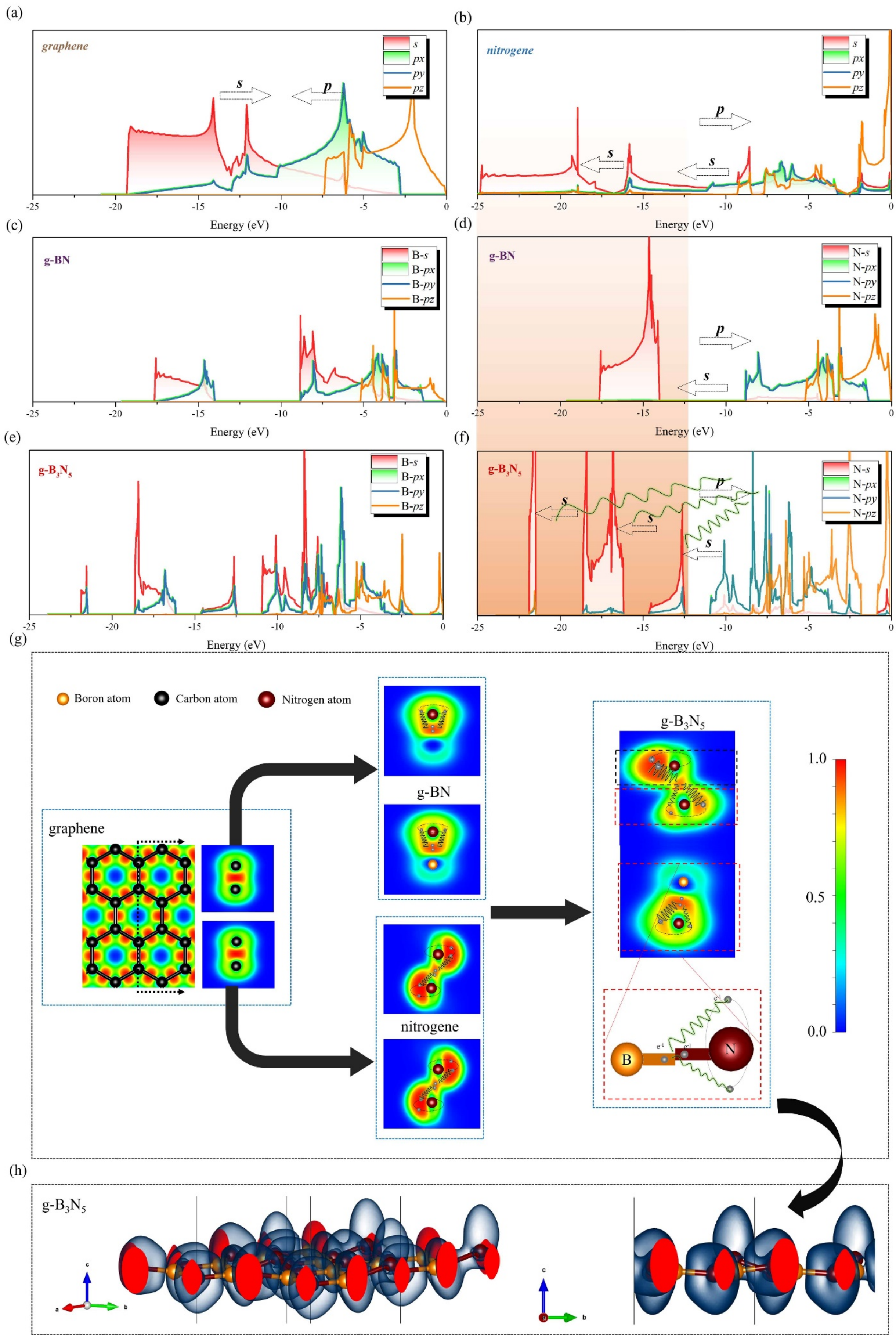

Figure 3. Comparative analysis of the partial density of states (*p*DOS) of (a) graphene, (b) nitrogene (c, d) g-BN, and (e, h) g-$B_3N_5$. (g) The evolution of the 2D electron localization function (ELF) with the structural reorganization. (f) The 3D ELF of g-$B_3N_5$ with the isosurface of 0.6.

## 2.4 Phonon-phonon scattering

At room temperature (300K)，the mode-level phonon properties are plotted in Fig. 4(a-d), including group velocity, relaxation time, scattering phase space, and Grüneisen parameter. As show in Fig. 4(a), the low in-plane acoustic group velocity comes from softened bond strength, *i.e.*, $C\text{-}C_{\text{bond}} > B\text{-}N_{\text{bond}} > N\text{-}N_{\text{bond}}$. Correspondingly, the smaller Young's modulus $E$ is also captured, as shown in Note S2 and Table S2. Moreover, as show in Fig. 4(b), the relaxation time of g-$B_3N_5$ is the lowest among the graphene, g-BN, nitrogene, g-$B_3N_5$. Thus the suppression of relaxation time is also responsible for the decreased $\kappa$ of g-$B_3N_5$, which is due to large scattering phase space [Fig. 4(c)] and large Grüneisen parameter Fig. 4(d). More information can be found in the Supplementary NoteS3.

Further insight into the anharmonicity can be obtained through the scattering channel and the scattering process, as shown in Fig. 4(e-l)[19,23]. For the planar structure, the phonon scattering of the FA branch of graphene and g-BN only involves the even-numbered FA phonons due to the phonon-phonon scattering selection rule based on symmetry[52], *i.e.*, FA+FA→TA/LA. Differently, the odd number of FA mode scattering channels of FA+FA→FA and FA→FA+FA are allowed in nitrogene as show in Fig. 5(e) owing to the buckling structure breaks the selection rule like silicene[19]. Intriguingly, the scattering channels involving the odd and even FA modes are simultaneously allowed in g-$B_3N_5$. The unique scattering channels can be attributed to the reorganization of their components. More scattering channels means more probability of scattering.

Moreover, the absorption process mainly contributes the scattering channels of the FA phonon modes in graphene and g-BN. Correspondingly, the scattering rate of the absorption process is higher than that of the emission process as shown in Fig. 4(i) and (j). However, the contribution of the absorption process in nitrogene and g-$B_3N_5$ is considerably reduced, and the emission process plays an essential role. Simultaneously, the normal (N) process scattering of g-$B_3N_5$ is suppressed dramatically, while the proportion of the N process in graphene is significantly higher than that of other materials. Weak N-process scattering means Weak phonon hydrodynamics[53], which indicates lower $\kappa$. Therefore, after recombination, the scattering channel increases, the absorption process and the N process are inhibited, resulting in a thermal conductivity of g-$B_3N_5$ that is two orders of magnitude lower than that of graphene. Thus, after recombination, the bond is softened, the scattering channel is increased, and the scattering of the absorption process and the N process is suppressed, leading to a $\kappa$ of g-$B_3N_5$ that is two orders of magnitude lower than that of graphene.

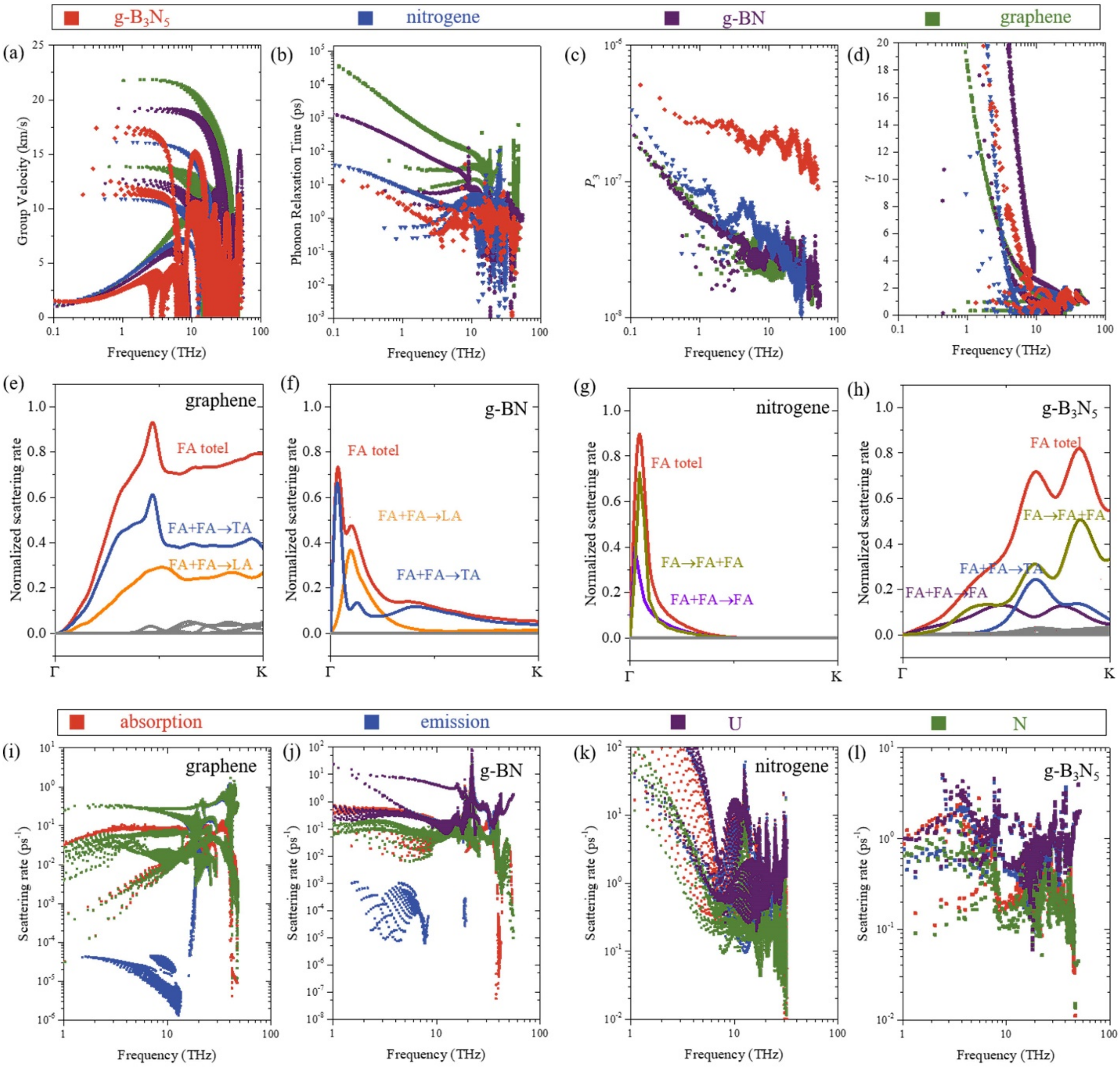

Figure 4. Comparative analysis of phonon transport properties at 300 K, including (a) Group velocity, (b) Relaxation time, (c) Phase space $P_3$, and (d) Grüneisen parameter (γ). The scattering channel of the FA branch of (e) graphene, (f) g-BN, (g) nitrogene, and (h) g-$B_3N_5$. Only the prominent channels are presented. Mode-level scattering process at 300 K for (i) graphene, (j) g-BN, (k) nitrogene, and (l) g-$B_3N_5$. Based on the conservation of energy, the three-phonon scattering process is split into the absorption process and the emission process, which is also split into the Normal process and the Umklapp process based on the conservation of momentum. The “absorption”, “emission”, “N” and "U" represent the absorption process, the emission process, the normal process, and the U process, respectively.

## 2.5 Thermal transport properties

Based on phonon properties, thermal conductivity of graphene, g-BN, nitrogene, and g-$B_3N_5$ can be obtained[54]. Fig. 5(a) shows the mode level cumulative $\kappa$ of four 2D materials at room temperature. In the entire frequency range, the $\kappa$ always maintains the relationship of graphene > g-BN > nitrogene > g-$B_3N_5$. The $\kappa$ of g-$B_3N_5$ from low-frequency mode is significantly suppressed due to the softening of the low-frequency acoustic phonon branches, which enhances the phonon-phonon scattering. Thus, the iterative $\kappa$ of g-$B_3N_5$ (21.08 W/mK) is two orders of magnitude lower than graphene, while g-BN (300.15W/mK) and nitrogene (101.50W/mK) are only one order of magnitude lower. Because of the low percentage of 13.89% (graphene), 27.23% (g-BN), and 32.79% (nitrogene), the relaxation time approximation (RTA) method will seriously underestimate $\kappa$. However, in g-$B_3N_5$, the $\kappa$ of RTA accounts for 91.07%, which implies the phonon hydrodynamics is broken down.

Furthermore, the percentage contribution of different phonon branches to the $\kappa$ is plotted in Fig. 3(c). The FA branch dominates the $\kappa$ of graphene, accounting for 84%, which is consistent with the previous result[51]. To reduce $\kappa$, it is necessary first to reduce the FA branch contribution, and then reduce the contribution of TA and LA branches. Therefore, due to the softening of the acoustic phonon branch, the $\kappa$ of g-$B_3N_5$ is significantly suppressed based on the recombination strategy. Furthermore, this strategy can be used to construct g-$B_3P_5$ and g-$B_3As_5$, and they exhibit similar geometry and band structure to g-$B_3N_5$ as shown in Fig. S2 and Table S3. Their dynamic stability can be demonstrated by phonon dispersion in Supplementary Fig. S3. Noted that they have an order of magnitude lower $\kappa$ compared to g-BP and g-BAs, as shown in Fig. 5(d). The low thermal conductivity originates from the low group velocity and relaxation time driven by the large scattering phase space and the Grüneisen parameter as shown in Supplementary Fig. S4 and S5.

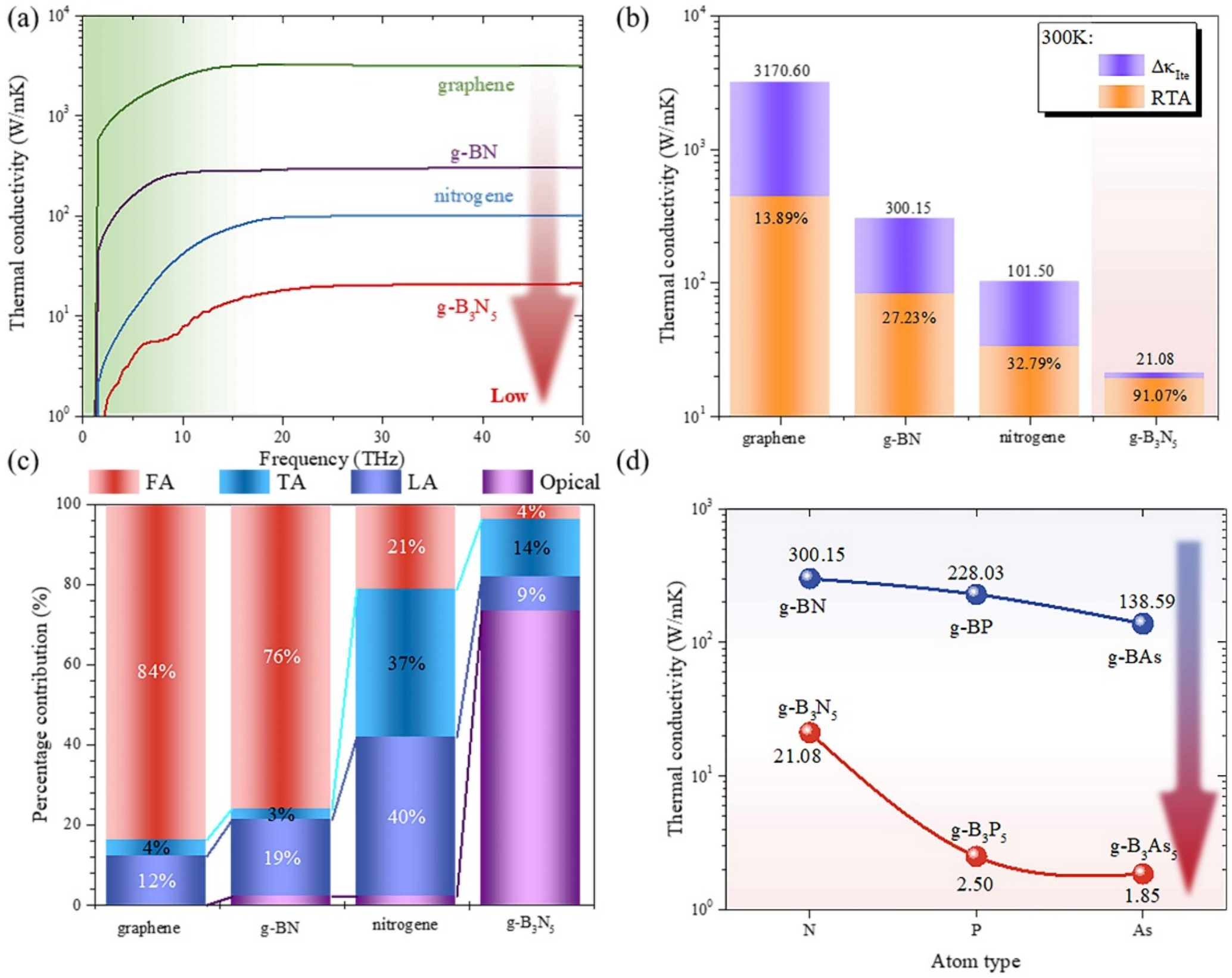

Figure 5. (a) The $\kappa$ of graphene, g-BN, nitrogene, and g-$B_3N_5$ as a function of frequency. (b) Comparison of the $\kappa$ calculated by the iterative (Ite) method and the relaxation time approximation (RTA) method at room temperature (300 K). Oranges indicate the results of the RTA method, and the percentages indicate the proportion of RTA in the iterative method. Purples represent the difference, *i.e.* $\Delta\kappa_{Ite} = \kappa_{Ite} - \kappa_{RTA}$. (c) The percentage contribution of each phonon branch to $\kappa$ at room temperature, including the flexural acoustic (FA), the transverse acoustic (TA), the longitudinal acoustic (LA), and the total optical (Optical) phonon branches. (d) Comparison of the $\kappa$ of g-$B_3X_5$ and g-BX (X=N, P, and As).

### 2.6 Enhanced effect of the strain regulation

Finally, we evaluate the synergy of electrons and geometric structures on thermal transport based on strain engineering. As shown in Fig. 6(a), the $\kappa$ of g-$B_3N_5$ increases with the strain increases until it approaches 2%, and then the $\kappa$ drops rapidly. The maximum $\kappa$ (69.35 W/mK) is obtained when the strain reaches 2%, which is 3.3 times higher than the pristine $\kappa$ (21.08 W/mK). Furthermore, to explain the enhancement, the group velocity and relaxation time of the mode levels at strains of 0%, 1%, and 2% are plotted in Fig. 6(b) and (c), respectively. The relaxation time of g-$B_3N_5$ decreases greatly while the group velocity does not change much, which reveals the significance of anharmonicity in the strain regulation.

The weakening of the anharmonicity comes from the reduction of the Grüneisen parameter, as shown in Fig. 6(d). Noted that the strain regulation effect for g-$B_3N_5$ is stronger than that for g-BN[55]. The $\kappa$ of g-BN increases with the strain up to 6%, which comes from the weakened anharmonicity between the lone pair electrons and the pair of bonded electrons. However, comparing to the 3.3 times enhancement in g-$B_3N_5$, it increases by less than twice in g-BN when the strain reaches 6%[55]. Thus, the lone pair of electrons is not the only protagonist in g-$B_3N_5$, and the buckling structure will also produce a specific response to strain. Based on above analysis, the synergy effect will enhance the regulatory effect of strain for g-$B_3N_5$.

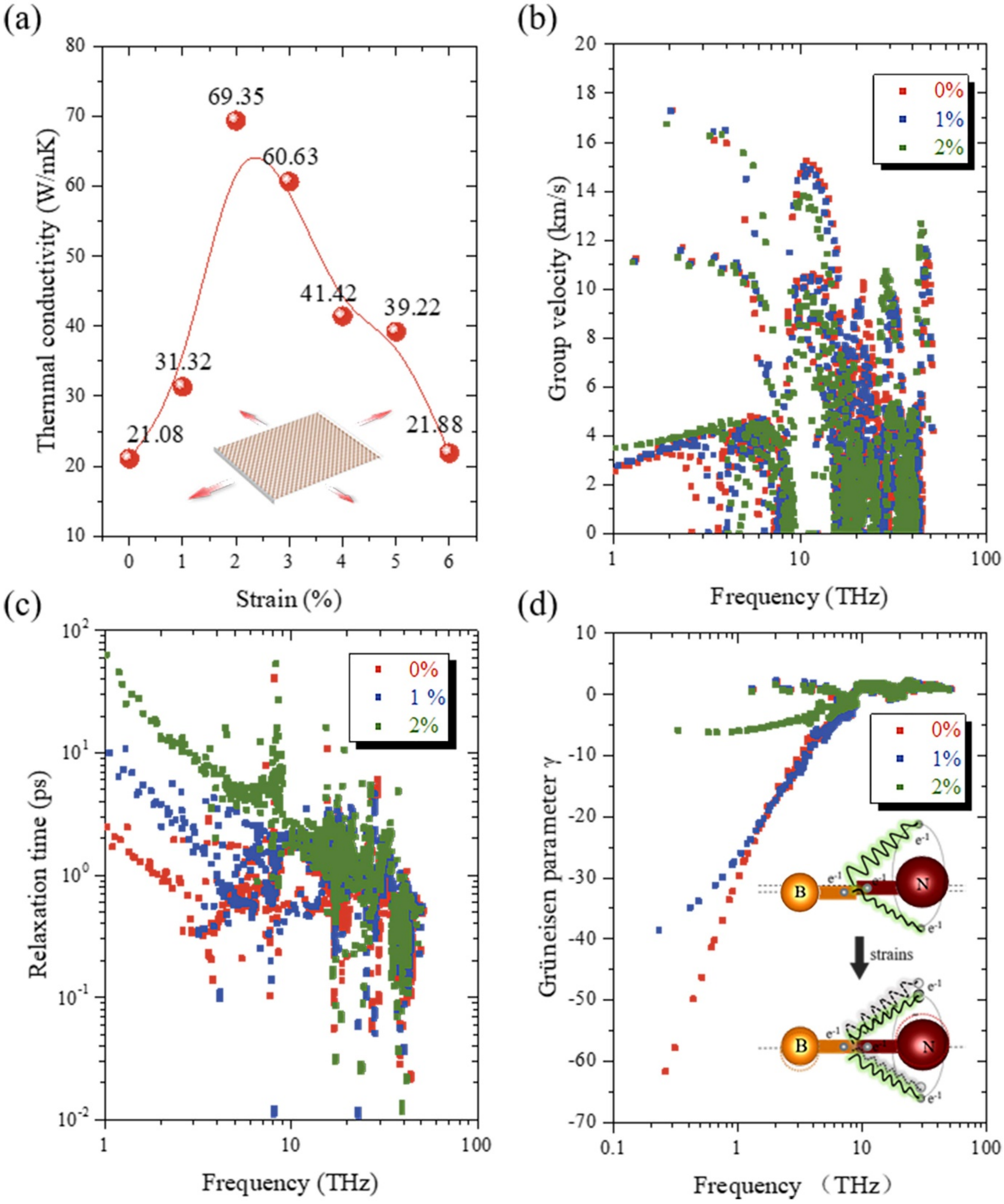

Figure 6. (a) The strain modulated $\kappa$ of g-$B_3N_5$. The insert is a schematic diagram of the applied tensile strain. Comparative analysis of (b) group velocity, (c) relaxation time, and (d) Grüneisen

parameter at 0%, 1% and 2% strain. The insert of (d) is a schematic diagram of the interactions between lone pair electrons and bonding electrons.

## 3. Conclusion

In summary, we propose a component reconstruction strategy to design a novel group of borides, *e.g.* g-$B_3N_5$. The $\kappa$ of g-$B_3N_5$ is 21.08 W/mK, which is two orders of magnitude lower than graphene. The origin of the low $\kappa$ can be is attributed to the synergy effect of electronics and geometry on thermal transport obtained by comparing it with graphene, g-BN, and nitrogene. After component recombination of g-BN and nitrogene, the acoustic phonon mode of g-$B_3N_5$ is softened, and the scattering process, including absorption process and the U process are suppressed. Furthermore, the asymmetry and unevenness of the local charge distribution in the system enhance the anharmonic force interaction. With this synergy strategy, we further designed g-$B_3P_5$ (2.50 W/mK) and g-$B_3As_5$ (1.85 W/mK) with ultra-low $\kappa$ based on the component reorganization strategy. Our research provides new insights for materials informatics in the design and search of low $\kappa$ materials. The g-$B_3N_5$ with low $\kappa$ expands the application of BN compounds in low $\kappa$ as a supplement to g-BN.

## 4. Computational Methodology

All first-principles calculations are performed by the *Vienna ab initio simulation package* (VASP)[56] with the Perdew–Burke–Ernzerhof (PBE)[57] functional. The kinetic energy cutoff of 1000eV is used to expand the wave functions for a plane-wave basis. A Monkhorst-Pack[58] ***q***-mesh of 15×15×1 (2×2×1) is used for structural optimization (force constant calculation) until the energy accuracy of $10^{-6}$ eV. The convergence accuracy of the Hellmann-Feynman force is set to $10^{-5}$ eV/Å. In addition, the force constant is calculated using the supercell of 3×3×1 for g-$B_3N_5$. To obtain the $\kappa$, the second-order and the third-order force constant are processed by the *PHONOPY*[59] and *thirdorder.py*[54] codes, respectively and input into the *ShengBTE* software[54].

## Acknowledgments

G.Q. is supported by the National Natural Science Foundation of China (Grant No. 52006057), the Fundamental Research Funds for the Central Universities (Grant Nos. 531119200237 and 541109010001), and the State Key Laboratory of Advanced Design and Manufacturing for Vehicle Body at Hunan University (Grant No. 52175011). X.Z. is supported by the Fundamental Research Funds for the Central Universities (Grant No. 531118010490) and the National Natural Science Foundation of China (Grant No. 52006059). H.W. is supported by the National Natural Science Foundation of China (Grant No. 51906097). The numerical calculations in this paper have been done on the supercomputing system of the National Supercomputing Center in Changsha. Simulations were performed with computing resources granted by RWTH Aachen University under project bund0011.

## References

1. Skoug, E. J. & Morelli, D. T. Role of lone-pair electrons in producing minimum thermal conductivity in nitrogen-group chalcogenide compounds. *Physical review letters* **107**, 235901 (2011).
2. Morelli, D. T., Jovovic, V. & Heremans, J. P. Intrinsically Minimal Thermal Conductivity in Cubic I − V − VI 2 Semiconductors. *Phys. Rev. Lett.* **101**, 035901 (2008).
3. Lee, S. *et al.* Resonant bonding leads to low lattice thermal conductivity. *Nat Commun* **5**, 3525 (2014).
4. Shportko, K. *et al.* Resonant bonding in crystalline phase-change materials. *Nature Mater* **7**, 653–658 (2008).
5. Qin, G. *et al.* Resonant bonding driven giant phonon anharmonicity and low thermal conductivity of phosphorene. *Phys. Rev. B* **94**, 165445 (2016).
6. Wu, J. *et al.* Systematic studies on anharmonicity of rattling phonons in type-I clathrates by low-temperature heat capacity measurements. *Phys. Rev. B* **89**, 214301 (2014).
7. Takabatake, T., Suekuni, K., Nakayama, T. & Kaneshita, E. Phonon-glass electron-crystal thermoelectric clathrates: Experiments and theory. *Reviews of Modern Physics* **86**, 669–716 (2014).
8. Pei, Y. *et al.* Multiple Converged Conduction Bands in $K_2Bi_8Se_{13}$: A Promising Thermoelectric Material with Extremely Low Thermal Conductivity. *J. Am. Chem. Soc.* **138**, 16364–16371 (2016).
9. Liu, Q. *et al.* Quasi-bonding driven abnormal isotropic thermal transport in intrinsically anisotropic nanostructure: a case of study of a phosphorus nanotube array. *Nanotechnology* **31**, 095704 (2019).
10. Shimoni-Livny, L., Glusker, J. P. & Bock, C. W. Lone Pair Functionality in Divalent Lead Compounds. *Inorg. Chem.* **37**, 1853–1867 (1998).
11. Egli, M. & Sarkhel, S. Lone Pair−Aromatic Interactions:  To Stabilize or Not to Stabilize. *Acc. Chem. Res.* **40**, 197–205 (2007).

12. Duncan, A. B. F. & Pople, J. A. The structure of some simple molecules with lone pair electrons. *Trans. Faraday Soc.* **49**, 217–224 (1953).

13. Shen, C. *et al.* Anomalously low thermal conductivity of two-dimensional GaP monolayers: A comparative study of the group GaX (X = N, P, As). *arXiv:2107.09964 [cond-mat]* (2021).

14. Nielsen, M. D., Ozolins, V. & Heremans, J. P. Lone pair electrons minimize lattice thermal conductivity. *Energy Environ. Sci.* **6**, 570–578 (2013).

15. Ok, K. M. *et al.* Bi-doped lanthanum molybdate: Enhancing the anharmonicity and reducing the thermal conductivity using Bi3+ with lone pair electrons. *Ceramics International* **44**, 15833–15838 (2018).

16. Dong, Y. *et al.* Bournonite $PbCuSbS_3$ : Stereochemically Active Lone-Pair Electrons that Induce Low Thermal Conductivity. *ChemPhysChem* **16**, 3264–3270 (2015).

17. Wang, X. & Liebau, F. Studies on bond and atomic valences. I. correlation between bond valence and bond angles in SbIII chalcogen compounds: the influence of lone-electron pairs. *Acta Crystallogr B Struct Sci* **52**, 7–15 (1996).

18. Wang, H. *et al.* Intrinsically low lattice thermal conductivity of monolayer hexagonal aluminum nitride (h-AlN) from first-principles: A comparative study with graphene. *International Journal of Thermal Sciences* **162**, 106772 (2021).

19. Qin, Z., Qin, G., Zuo, X., Xiong, Z. & Hu, M. Orbitally driven low thermal conductivity of monolayer gallium nitride (GaN) with planar honeycomb structure: a comparative study. *Nanoscale* **9**, 4295–4309 (2017).

20. Li, S. *et al.* Different Effects of Mg and Si Doping on the Thermal Transport of Gallium Nitride. *Frontiers in Materials* **8**, (2021).

21. Yu, L. *et al.* Abnormal enhancement of thermal conductivity by planar structure: A comparative study of graphene-like materials. *International Journal of Thermal Sciences* **174**, 107438 (2022).

22. Lindsay, L., Broido, D. A. & Reinecke, T. L. First-Principles Determination of Ultrahigh Thermal Conductivity of Boron Arsenide: A Competitor for Diamond? *Phys. Rev. Lett.* **111**, 025901 (2013).

23. Qin, G., Qin, Z., Wang, H. & Hu, M. Anomalously temperature-dependent thermal conductivity of monolayer GaN with large deviations from the traditional 1 / T law. *Phys. Rev. B* **95**, 195416 (2017).

24. Zhao, Y. *et al.* Porous architecture and thermal properties of thermal barrier coatings deposited by suspension plasma spray. *Surface and Coatings Technology* **386**, 125462 (2020).

25. Clarke, D. R., Oechsner, M. & Padture, N. P. Thermal-barrier coatings for more efficient gas-turbine engines. *MRS Bull.* **37**, 891–898 (2012).

26. Siegrist, T., Merkelbach, P. & Wuttig, M. Phase Change Materials: Challenges on the Path to a Universal Storage Device. *Annual Review of Condensed Matter Physics* **3**, 215–237 (2012).

27. Wuttig, M. & Yamada, N. Phase-change materials for rewriteable data storage. *Nature materials* **6**, 824–832 (2007).

28. Tolborg, K. & Iversen, B. B. Chemical Bonding Origin of the Thermoelectric Power Factor in Half-Heusler Semiconductors. *Chem. Mater.* acs.chemmater.1c01409 (2021) doi:10.1021/acs.chemmater.1c01409.

29. Agarwal, J., Sahoo, S., Mohanty, S. & Nayak, S. K. Progress of novel techniques for lightweight automobile applications through innovative eco-friendly composite materials: A review. *Journal of Thermoplastic Composite Materials* **33**, 978–1013 (2020).

30. Song, Z. & Zhao, X. Research on Lightweight Design of Automobile Lower Arm Based on Carbon Fiber Materials. *World Journal of Engineering and Technology* **5**, 730–742 (2017).

31. Zhu, L., Li, N. & Childs, P. R. N. Light-weighting in aerospace component and system design. *Propulsion and Power Research* **7**, 103–119 (2018).

32. Fetisov, K. V. & Maksimov, P. V. Topology optimization and laser additive manufacturing in design process of efficiency lightweight aerospace parts. *J. Phys.: Conf. Ser.* **1015**, 052006 (2018).

33. Hong, Y., Zhang, J., Huang, X. & Zeng, X. C. Thermal conductivity of a two-dimensional phosphorene sheet: a comparative study with graphene. *Nanoscale* **7**, 18716–18724 (2015).

34. Peng, B. *et al.* The conflicting role of buckled structure in phonon transport of 2D group-IV and group-V materials. *Nanoscale* **9**, 7397–7407 (2017).

35. Peng, B. *et al.* Phonon transport properties of two-dimensional group-IV materials from *ab initio* calculations. *Phys. Rev. B* **94**, 245420 (2016).

36. Jo, I. *et al.* Thermal Conductivity and Phonon Transport in Suspended Few-Layer Hexagonal Boron Nitride. *Nano Lett.* **13**, 550–554 (2013).

37. Wang, C. *et al.* Superior thermal conductivity in suspended bilayer hexagonal boron nitride. *Sci Rep* **6**, 25334 (2016).

38. Lindsay, L. & Broido, D. A. Enhanced thermal conductivity and isotope effect in single-layer hexagonal boron nitride. *Phys. Rev. B* **84**, 155421 (2011).

39. Zhou, H. *et al.* High thermal conductivity of suspended few-layer hexagonal boron nitride sheets. *Nano Res.* **7**, 1232–1240 (2014).

40. Lindsay, L. & Broido, D. A. Theory of thermal transport in multilayer hexagonal boron nitride and nanotubes. *Phys. Rev. B* **85**, 035436 (2012).

41. Taheri, A., Da Silva, C. & Amon, C. H. Phonon thermal transport in β- N X (X= P, As, Sb) monolayers: A first-principles study of the interplay between harmonic and anharmonic phonon properties. *Physical Review B* **99**, 235425 (2019).

42. Qin, G., Qin, Z., Wang, H. & Hu, M. Lone-pair electrons induced anomalous enhancement of thermal transport in strained planar two-dimensional materials. *Nano Energy* **50**, 425–430 (2018).

43. Xie, H., Hu, M. & Bao, H. Thermal conductivity of silicene from first-principles. *Appl. Phys. Lett.* **104**, 131906 (2014).

44. Zhang, X. *et al.* Thermal conductivity of silicene calculated using an optimized Stillinger-Weber potential. *Phys. Rev. B* **89**, 054310 (2014).

45. Wang, H. *et al.* The exceptionally high thermal conductivity after 'alloying' two-dimensional gallium nitride (GaN) and aluminum nitride (AlN). *Nanotechnology* **32**, 135401 (2021).

46. Mortazavi, B. *et al.* Outstanding strength, optical characteristics and thermal conductivity of graphene-like BC3 and BC6N semiconductors. *Carbon* **149**, 733–742 (2019).

47. Wang, H., Qin, G., Li, G., Wang, Q. & Hu, M. Low thermal conductivity of monolayer ZnO and its anomalous temperature dependence. *Phys. Chem. Chem. Phys.* **19**, 12882–12889 (2017).

48. Delaire, O. *et al.* Giant anharmonic phonon scattering in PbTe. *Nature Mater* **10**, 614–619 (2011).

49. Choudhry, U., Yue, S. & Liao, B. Origins of significant reduction of lattice thermal conductivity in graphene allotropes. *Phys. Rev. B* **100**, 165401 (2019).

50. Gan, Y., Huang, Y., Miao, N., Zhou, J. & Sun, Z. Novel IV–V–VI semiconductors with ultralow lattice thermal conductivity. *J. Mater. Chem. C* **9**, 4189–4199 (2021).

51. Gu, X. & Yang, R. First-principles prediction of phononic thermal conductivity of silicene: A comparison with graphene. *Journal of Applied Physics* **117**, 025102 (2015).

52. Lindsay, L., Broido, D. A. & Mingo, N. Flexural phonons and thermal transport in graphene. *Phys. Rev. B* **82**, 115427 (2010).

53. Lee, S., Broido, D., Esfarjani, K. & Chen, G. Hydrodynamic phonon transport in suspended graphene. *Nat Commun* **6**, 6290 (2015).

54. Li, W., Carrete, J., A. Katcho, N. & Mingo, N. ShengBTE: A solver of the Boltzmann transport equation for phonons. *Computer Physics Communications* **185**, 1747–1758 (2014).

55. Qin, G., Qin, Z., Wang, H. & Hu, M. Lone-pair electrons induced anomalous enhancement of thermal transport in strained planar two-dimensional materials. *Nano Energy* **50**, 425–430 (2018).

56. Kresse, G. & Hafner, J. *Ab initio* molecular-dynamics simulation of the liquid-metal–amorphous-semiconductor transition in germanium. *Phys. Rev. B* **49**, 14251–14269 (1994).

57. Perdew, J. P., Burke, K. & Ernzerhof, M. Generalized Gradient Approximation Made Simple. *Phys. Rev. Lett.* **77**, 3865–3868 (1996).

58. Monkhorst, H. J. & Pack, J. D. Special points for Brillouin-zone integrations. *Phys. Rev. B* **13**, 5188–5192 (1976).

59. Chaput, L., Togo, A., Tanaka, I. & Hug, G. Phonon-phonon interactions in transition metals. *Phys. Rev. B* **84**, 094302 (2011).

# Supplemental material

## Note S1 Comparative analysis of phonon dispersion

The phonon dispersion and phonon density of states (Dos) of graphene, g-BN, nitrogene, and g-$B_3N_5$ along the high symmetry point Γ-M-K-Γ are plotted in Fig. S1 (a-d), respectively. The absence of imaginary frequencies indicates the stability of the dynamics of the material. Graphene, g-BN, and nitrogene have six branches, including three acoustic branches (FA, TA, LA) and three optical branches (FO, TO, LO), which originated from only two atoms in the primitive cell. In addition to three acoustic branches, g-$B_3N_5$ has twenty-one optical branches due to the eight atoms in the primitive cell. The FA branch of g-$B_3N_5$ is similar to graphene, g-BN and nitrogene, all of which are flexural and have a quadratic relationship when the TA and LA branch show a linear relationship, which is a characteristic of the typical dispersion relationship for two-dimensional materials[3,4,8–10]. The phonon band gap of g-$B_3N_5$ (0.6 and 0.9 THz) is extremely weak and similar to g-BN (0.9 THz), which is due to the similar atomic mass between the boron (B) atom and the nitrogen (N) atom (B ~ 10.8 and N ~ 14.0) when graphene and nitrogene do not have a phonon band gap due to the same atomic composition.

## Note S2 Assessment of atomic bonding

To explain the ultra-low thermal conductivity of g-$B_3N_5$, a mode-level analysis of the phonon properties is shown in Fig 4. Compared with graphene, the acoustic group velocities of g-BN, nitrogene, and g-$B_3N_5$ are significantly suppressed in Fig. 4(a), which is due to the softened acoustic phonon branches. The decrease in group velocity can be further evaluated through specific mechanical properties, which can intuitively reflect the weakening of bonding strength. The sound velocity of TA ($\upsilon_T$) branch and LA ($\upsilon_L$) branch can be measured[11]:

$$\upsilon_L = \sqrt{\frac{K+G}{\rho_{2D}}},$$

$$\upsilon_T = \sqrt{\frac{G}{\rho_{2D}}}$$

where $\kappa$, $\mu$, and $\rho_{2D}$ are in-plane stiffness, the shear modulus, and density. $\kappa$, and $\mu$ can be obtained by[11]:

$$K = \frac{E_{2D}}{2(1-\upsilon)},$$

$$G = \frac{E_{2D}}{(1+\upsilon)2},$$

where $E_{2D}$, $\upsilon$ are the in-plane Young's modulus and Poisson's ratio, respectively. The results of $\upsilon_T$ and $\upsilon_T$ are listed in Table S2, which can fit well with the results of the mode-level group velocity.

## S3 Analysis of phonon properties

As shown in Fig. 4(c), the largest scattering phase space appears in g-$B_3N_5$, even larger than nitrogene, while the lowest scattering phase space appears in graphene. The large phase space is due to the uneven buckling structure. The scattering phase space of g-BN is almost the same as that of graphene due to the same planar structure. The scattering phase space of nitrogene is enhanced, as the buckling structure breaks the plane symmetry. Noted that the buckling height of g-$B_3N_5$ is lower than that of nitrogene, but the scattering phase space is significantly enhanced. The large enhancement of the scattering phase space is due to the combination of the local plane and the buckling structure formed in the primitive cell, which makes the atomic volume more unevenly distributed in the space. The uneven distribution of atoms is intuitively reflected in the diversity of bond lengths and bond angles. The larger scattering phase space provides more opportunities for phonon-phonon scattering by increasing the number of scattering channels, thereby reducing thermal conductivity. Furthermore, the larger Grüneisen parameter $\gamma$ can be captured in g-$B_3N_5$ as shown in Fig. 4(d). The largest Grüneisen parameter is found in g-BN, which is due to the lone pair of electrons excited by the polarization in the nitrogen atom. With the formation of N-N bonds and the reduction of B-N bonds, the polarization is weakened in g-$B_3N_5$. Thus, the Grüneisen parameter of g-$B_3N_5$ is smaller than g-BN, but it is still stronger than graphene and nitrogene. Ultimately, the large scattering phase space and Grüneisen parameter lead to a low relaxation time for g-$B_3N_5$. Similar phonon properties are also shown in g-$B_3P_5$ [Fig. S4] and g-$B_3As_5$ [Fig. S5].

Table S1. Thermal conductivity of graphene-like materials.

| Thermal conductivity (W/mK) | Average atomic mass | Type | Ref |
|---|---|---|---|
| 410.00 | 11.71 | g-$C_3B$ | Ref[1] |
| 3197.00 | 12.01 | graphene | This study |
| 1080.00 | 12.11 | g-$BC_6N$-1 | Ref [1] |
| 1430.00 | 12.11 | g-$BC_6N$-2 (aimchair) | Ref [1] |
| 1710.00 | 12.11 | g-$BC_6N$-2 (zigzag) | Ref [1] |
| 300.15 | 12.41 | g-BN | This study |
| 103.02 | 12.5 | g-$C_3N$ | Ref [2] |
| 21.08 | 12.81 | g-$B_3N_5$ | This study |
| 101.50 | 14.00 | nitrogene | This study |
| 74.42 | 20.49 | g-AlN | Ref [3] |
| 28.62 | 22.49 | *g-NP* | Ref [4] |
| 29.57 | 26.2 | $g\text{-}Ga_{0.25}Al_{0.75}N$ | Ref [5] |
| 19.00 | 28.09 | *silicene* | Ref [6] |
| 23.00 | 30.97 | phorphosene | Ref [4] |
| 21.49 | 31.18 | $g\text{-}Ga_{0.5}Al_{0.5}N$ | Ref [5] |
| 11.90 | 36.52 | $g\text{-}Ga_{0.75}Al_{0.25}N$ | Ref [5] |
| 4.50 | 40.7 | g-ZnO | Ref [7] |
| 14.92 | 41.86 | g-GaN | Ref [8] |
| 38.46 | 44.46 | g-NAs | Ref [4] |
| 7.07 | 74.92 | As | Ref [6] |
| 34.66 | 67.88 | g-NSb | Ref [4] |
| 1.02 | 121.76 | Sb | Ref [6] |

Table S2. Comparison of Poisson's ratio $\upsilon$, elastic modulus $E_{2D}$, in-plane stiffness $K$, shear modulus $G$, sound velocity of TA branch $\upsilon_T$ and LA branch $\upsilon_L$.

| | graphene | g-BN | nitrogene | g-$B_3N_5$ |
|---|---|---|---|---|
| $v$ | 0.22 | 0.22 | 0.10 | 0.11 |
| $E_{2D}$ (N/m) | 328.70 | 278.62 | 267.40 | 244.77 |
| $K$ (N/m) | 210.44 | 179.52 | 147.90 | 137.05 |
| $G$ (N/m) | 134.82 | 113.81 | 122.00 | 110.56 |
| $\upsilon_T$ (N/m) | 21.37 | 19.73 | 16.28 | 17.48 |
| $\upsilon_L$ (N/m) | 13.35 | 12.29 | 10.94 | 11.68 |

Table S3. Geometric parameters of g-$B_3X_5$ (X =N, P, As).

| Type | g-$B_3N_5$ | g-$B_3P_5$ | g-$B_3As_5$ |
|---|---|---|---|
| *a* | 4.92 | 6.37 | 6.75 |
| *h* | 0.50 | 1.20 | 1.48 |

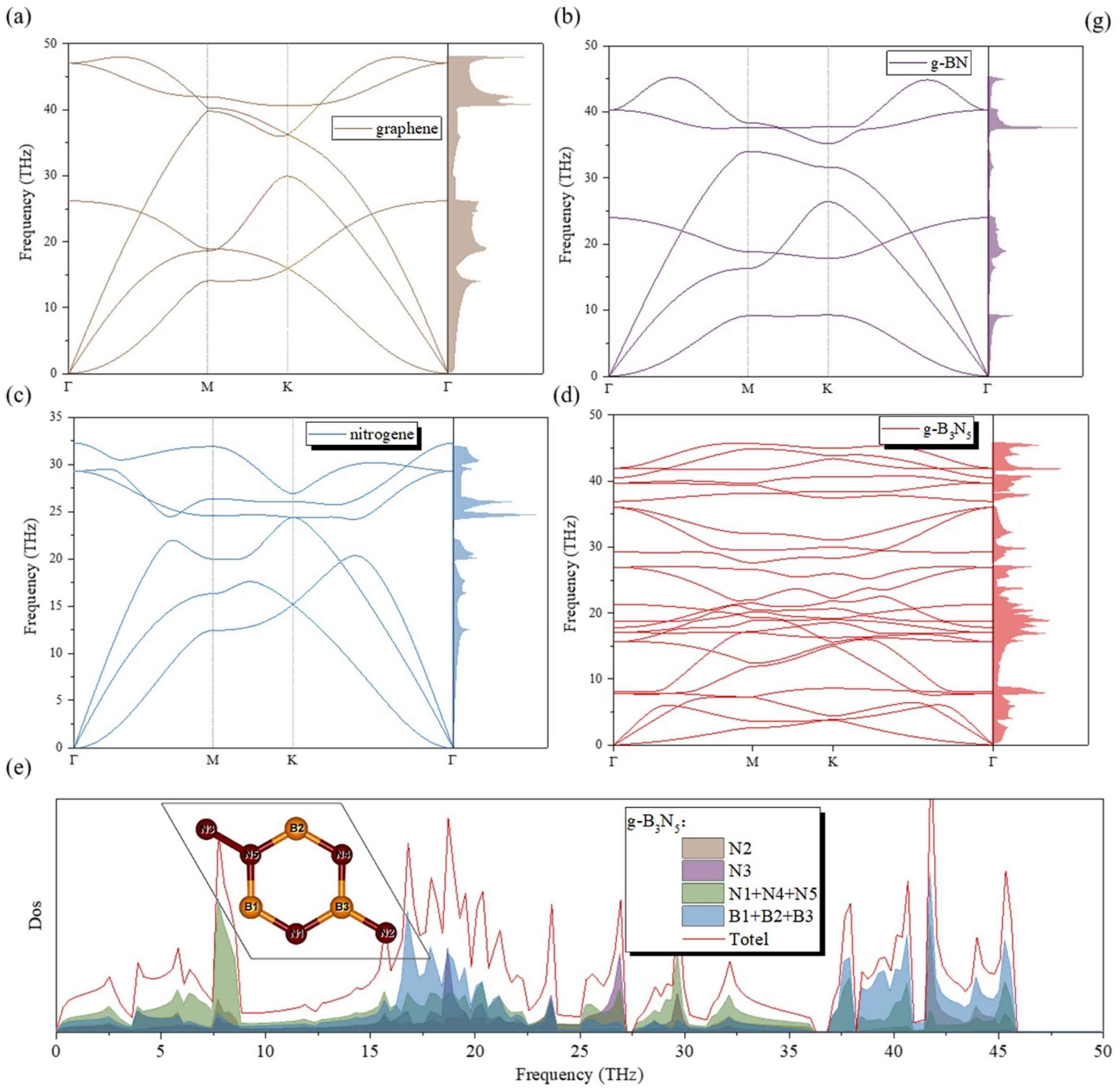

Figure S1. Comparative analysis of phonon dispersion of (a) graphene, (b) g-BN, (c) nitrogene and (d) g-$B_3N_5$. (e) Analysis of Phonon Density of State of g-$B_3N_5$.

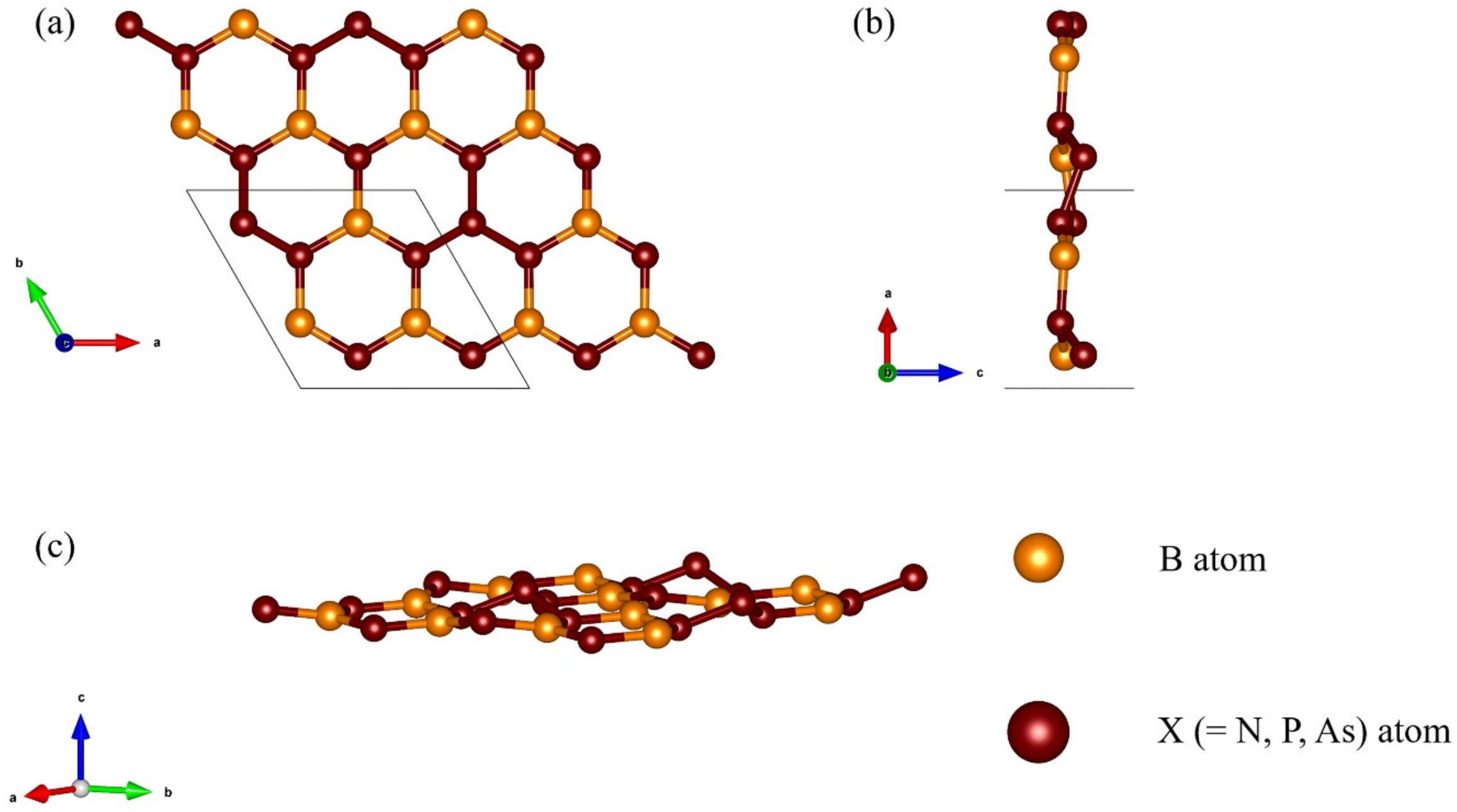

Figure S2. Geometry structure of g-$B_3X_5$ (X =N, P, and As).

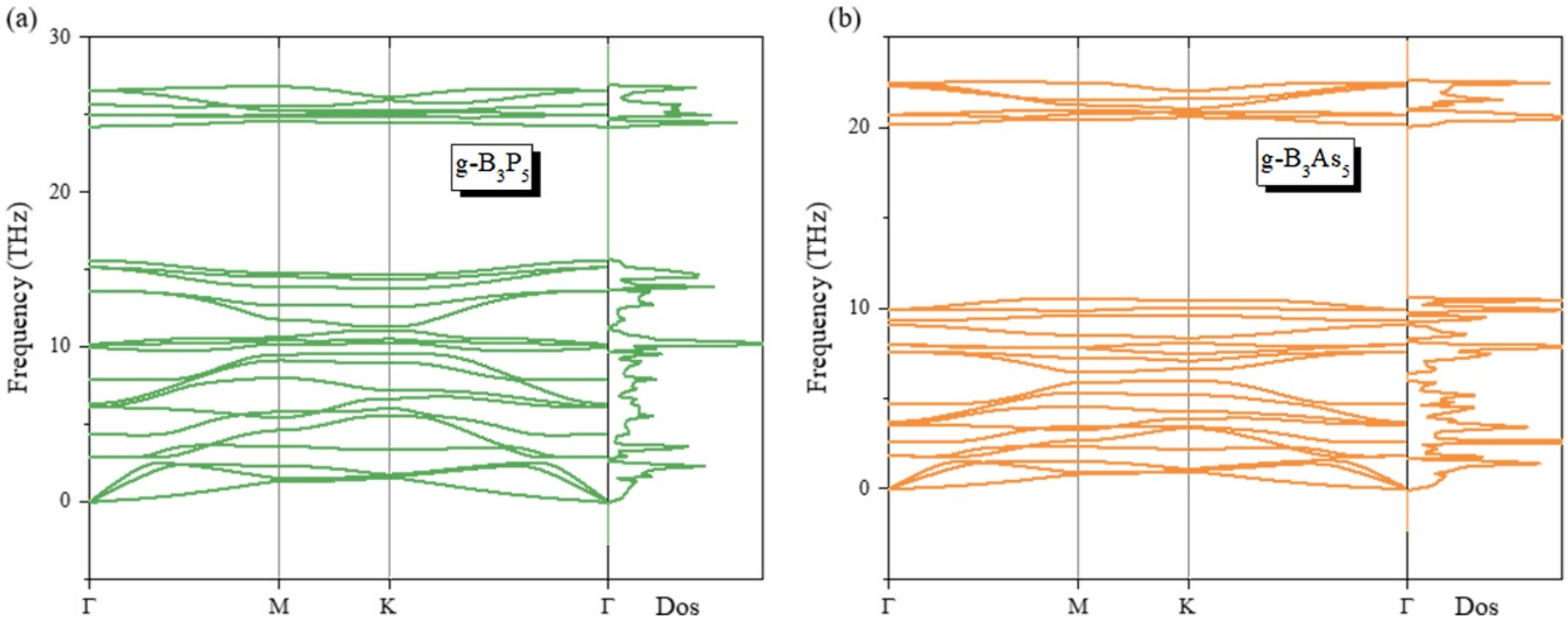

Figure S3. The phonon dispersion and phonon density of states of (a) g-$B_3P_5$ and (b) g-$B_3As_5$, where there is no imaginary frequency means stability.

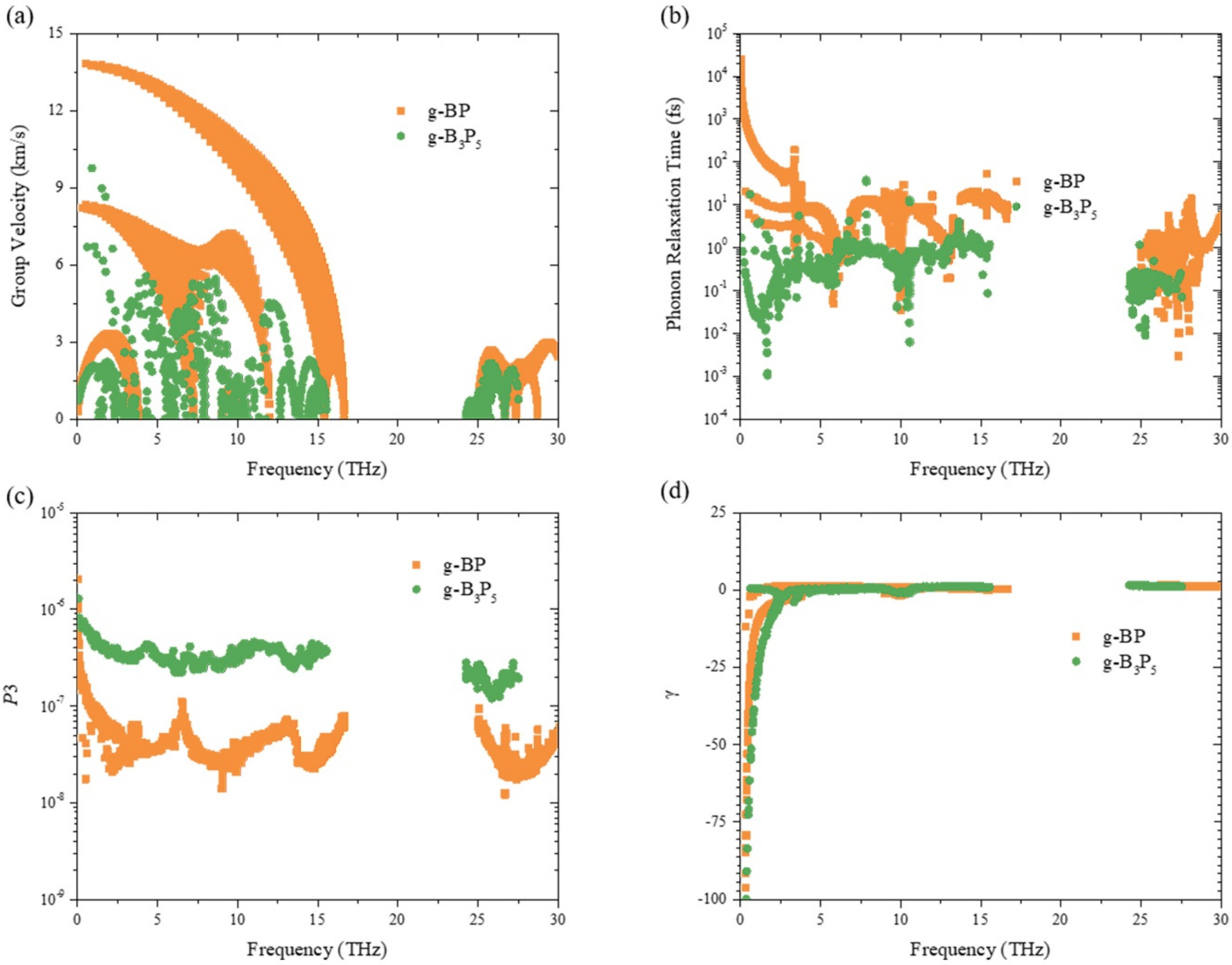

Figure S4. Comparative analysis of phonon properties between g-BP and g-$B_3P_5$. (a) The group velocity, (b) phonon relaxation time, (c) phase space and (d) Grüneisen parameter.

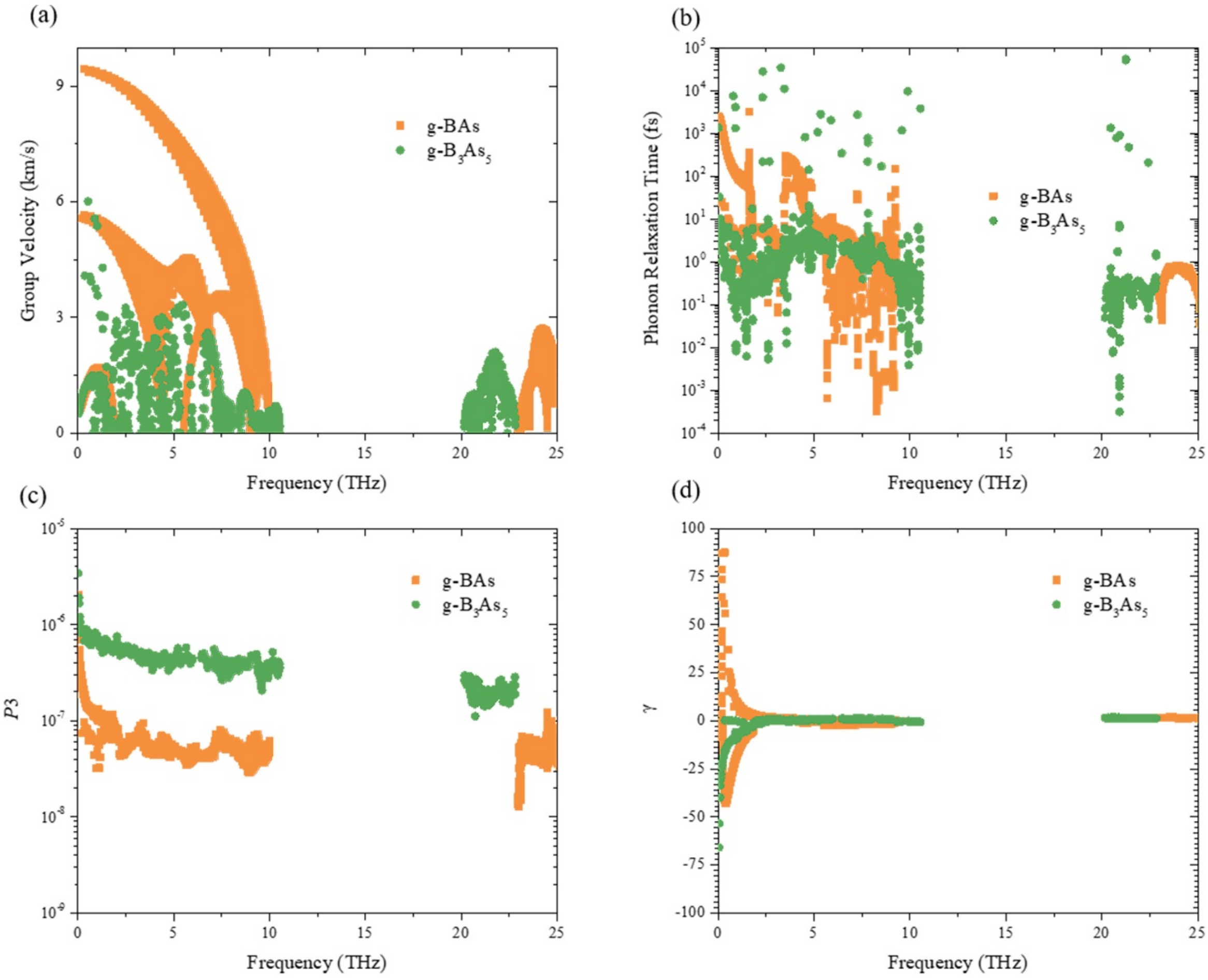

Figure S5. Comparative analysis of phonon properties between g-BAs and g-$B_3As_5$. (a) The group velocity, (b) phonon relaxation time, (c) phase space and (d) Grüneisen parameter.

## Reference

1. Mortazavi, B. *et al.* Outstanding strength, optical characteristics and thermal conductivity of graphene-like BC3 and BC6N semiconductors. *Carbon* **149**, 733–742 (2019).
2. Wang, H. *et al.* Lone-Pair Electrons Do Not Necessarily Lead to Low Lattice Thermal Conductivity: An Exception of Two-Dimensional Penta-CN $_2$. *J. Phys. Chem. Lett.* **9**, 2474–2483 (2018).
3. Wang, H. *et al.* Intrinsically low lattice thermal conductivity of monolayer hexagonal aluminum nitride (h-AlN) from first-principles: A comparative study with graphene. *International Journal of Thermal Sciences* **162**, 106772 (2021).
4. Taheri, A., Da Silva, C. & Amon, C. H. Phonon thermal transport in β- N X (X= P, As, Sb) monolayers: A first-principles study of the interplay between harmonic and anharmonic phonon properties. *Physical Review B* **99**, 235425 (2019).
5. Wang, H. *et al.* The exceptionally high thermal conductivity after 'alloying' two-dimensional gallium nitride (GaN) and aluminum nitride (AlN). *Nanotechnology* **32**, 135401 (2021).
6. Peng, B. *et al.* The conflicting role of buckled structure in phonon transport of 2D group-IV and group-V materials. *Nanoscale* **9**, 7397–7407 (2017).
7. Wang, H., Qin, G., Li, G., Wang, Q. & Hu, M. Low thermal conductivity of monolayer ZnO and its anomalous temperature dependence. *Phys. Chem. Chem. Phys.* **19**, 12882–12889 (2017).
8. Qin, G., Qin, Z., Wang, H. & Hu, M. Anomalously temperature-dependent thermal conductivity of monolayer GaN with large deviations from the traditional 1 / T law. *Phys. Rev. B* **95**, 195416 (2017).

9. Hong, Y., Zhang, J. & Zeng, X. C. Monolayer and bilayer polyaniline $C_3N$: two-dimensional semiconductors with high thermal conductivity. *Nanoscale* **10**, 4301–4310 (2018).

10. Lin, W., Liang, S.-D., Li, J. & Yao, D.-X. Phonon dispersions and electronic structures of two-dimensional IV-V compounds. *Carbon* **172**, 345–352 (2021).

11. Politano, A. & Chiarello, G. Probing the Young's modulus and Poisson's ratio in graphene/metal interfaces and graphite: a comparative study. *Nano Res.* **8**, 1847–1856 (2015).